\documentclass[showpacs,floatfix,preprintnumbers,eqsecnum,aps,nofootinbib]{revtex4} 

\usepackage{epsfig}
\usepackage{citesort}

\def \bsllg     {$B_s \to \ell^+ \ell^- \gamma$ }
\def \bsmupmg   {$B_s \to \mu^+ \mu^- \gamma$ }
\def \bstaupmg  {$B_s \to \tau^+ \tau^- \gamma$ }
\def \bsll      {$B_s \to \ell^+ \ell^-$ }
\def \btosllg   {$b \to s \ell^+ \ell^- \gamma$ }
\def \btosll    {$b \to s \ell^+ \ell^-$ }
\def \btoxsga   {$B \to X_s \gamma$ }

\def \cq#1      {C_{Q_#1} }
\def \c#1eff    {C_#1^{eff} }

\def \pl        {{\rm $P_L$ }}
\def \pn        {{\rm $P_N$ }}
\def \pt        {{\rm $P_T$ }}
\def \tanbeta   {$tan\beta $ } 
\def \epsmualbesig  {\varepsilon_{\mu\alpha\beta\sigma} }
\def \modmsq    {|{\cal M}|^2 }
\def \modm#1sq  {|{\cal M}_#1|^2 }
\def \rem12     {Re({\cal M}_1 {\cal M}^*_2) }
\def \facmatrix {| \frac{\alpha^{3/2} G_F}{\sqrt{2 \pi}} V_{tb}
                 V_{ts}^*|^2 } 
\def \modsq#1   {|#1|^2}
\def \rea#1#2   {Re(#1^* #2)}
\def \facdr     {| \frac{\alpha^{3/2} G_F}{2 \sqrt{2 \pi}} V_{tb}
                 V_{ts}^*|^2 } 
\def \ctencqtwo {(C_{10} + \frac{m_{B_s}^2}{2 m_\ell m_b} C_{Q_2})}
\def \drcqone   {(\frac{m_{B_s}^2}{2 m_\ell m_b} C_{Q_1})}
\def \lnzh      {{\rm ln}(\hat{z})}

\def \ml        {$m_\ell$ }
\def \mle#1     {m_\ell^#1 }
\def \mbs       {m_{B_s}}
\def \mb#1      {m_{B_s}^#1}

\def \psq       {p^2}
\def \p#1qsq    {(p_#1 q)^2}
\def \pqsq      {(p q)^2}

\def \sh        {\hat{s}}
\def \mlh       {\hat{m}_\ell}
\def \mlhsq     {\hat{m}_\ell^2}
\def \zh        {\hat{z}}
\def \faco      {\sqrt{1 - \frac{4 \hat{m}_\ell^2}{\hat{s}}}}
\def \fact      {(1 - \frac{4 \hat{m}_\ell}{\hat{s}})}

\def \pv        {{\bf p}_1}
\def \qv        {{\bf q}}
\def \ev#1      {{\bf e}_#1}

\def \beq       {\begin{equation}}
\def \eeq       {\end{equation}}
\def \beqa      {\begin{eqnarray}}
\def \eeqa      {\end{eqnarray}}
\def \bfig      {\begin{figure}}
\def \efig      {\end{figure}}
\def \bcen      {\begin{center}}
\def \ecen      {\end{center}}

\def \ie        {{\it i.e. }}
\def \eg        {{\it e.g. }}
\def \etal      {{\it et. al. }}

\def \prd#1#2#3       {Phys. \ Rev. {\bf D #1}, {#2} (#3)}
\def \pr#1#2#3        {Phys. \ Rev. {\bf #1}, {#2} (#3)}
\def \prl#1#2#3       {Phys. \ Rev. \ Lett. {\bf #1}, {#2} (#3)}
\def \nuclphysb#1#2#3 {Nucl. \ Phys. {\bf B #1}, {#2} (#3)}
\def \plb#1#2#3       {Phys. \ Lett. {\bf B #1}, {#2} (#3)}
\def \physrep#1#2#3   {Phys. \ Rep {\bf #1}, {#2} (#3)}
\def \zphysc#1#2#3    {Z. \ Phys. {\bf C #1}, {#2} (#3)}

\begin{document}

\renewcommand{\thefootnote}{\fnsymbol{footnote}}

\preprint{hep-ph/0203041}

\title{ Lepton polarization asymmetry in radiative dileptonic
B-meson decays in MSSM} 

\author{S. Rai Choudhury}
 \email{src@ducos.ernet.in}
\author{Naveen Gaur}
 \email{naveen@physics.du.ac.in} 
\author{Namit Mahajan}
 \email{nmahajan@physics.du.ac.in} 

\affiliation{ Department of Physics and Astrophysics, \\
     University of Delhi, \\
     Delhi - 110 007, India.}

\date{\today}

\pacs{13.20He,12.60.-i,13.88+e}


\begin{abstract}
\noindent In this paper we study the polarization asymmetries of the
final state lepton in the radiative dileptonic decay of B meson
(\bsllg) in the framework of Minimal Supersymmetric Standard Model
(MSSM) and various other unified models within the framework of MSSM
\eg mSUGRA, SUGRA (where condition of universality of scalar
masses is relaxed) etc.  Lepton polarization, in addition of having a 
longitudinal component (\pl), can have two other components, \pt and \pn,
lying in and perpendicular to the decay plane, which are proportional
to \ml and hence are significant for final state being $\mu^+ ~ \mu^-$
or $\tau^+ ~\tau^-$. We analyse the dependence of these
polarization asymmetries on the parameters of the various models. 
\end{abstract}

\maketitle





\section{\label{sectin:1}Introduction}

 Flavor Changing Neutral Current (FCNC) induced B-meson rare
decays provide a unique testing  
ground for Standard Model (SM) improved by QCD corrections via Operator
Product Expansion ( for a review and complete set of references see
\cite{Ali:1997vt}). Studies of rare B decays can give
 precise information about various fundamental parameters of SM like 
Cabbibo-Kobayashi-Maskawa (CKM) matrix elements, leptonic decay
constants etc. In addition to this, rare B decays can also give
information about various extensions of SM like two Higgs doublet 
model (2HDM)
\cite{Iltan:2000iw,Aliev:1999kr,Skiba:1993mg,Dai:1997vg} ,
Minimal Supersymmetric Standard Model (MSSM)
\cite{Cho:1996we,Hewett:1996ct,Choudhury:1999ze,RaiChoudhury:1999qb,Xiong:2001up,Huang:1999vb,goto1,Bobeth:2001jm}
etc. After the first observation of the penguin induced decay 
\btoxsga and the corresponding exclusive decay channel $B \to K^*
\gamma$ by CLEO \cite{cleo1} , rare decays have begun to play an
important role in particle physics phenomenology.  

\par Among the rare B decays,  \bsllg ($\ell = e, \mu, \tau$) are of  
special interest due to their relative cleanliness and sensitivity to
new physics. They have been extensively studied within SM 
\cite{Eilam:1997vg,Aliev:1997ud,Dincer:2001hu} and beyond
\cite{Iltan:2000iw}. In the mode \bsllg, one can study many
experimentally accessible 
quantities associated with final state leptons and photon \eg lepton
pair invariant mass spectrum, lepton pair forward backward asymmetry, 
photon energy distribution and various polarization asymmetries (like
longitudinal, transverse and normal). The final state leptons in the
radiative decay mode \bsllg, apart from having longitudinal
polarization, can have two more components of polarization ( \pt is
the component of the polarization lying in the decay plane and \pn is
the one that is normal to the decay plane)
\cite{Kruger:1996cv}. Both \pn and \pt remain non-trivial for $\mu^+
\mu^-$ and $\tau^+ \tau^-$ channel since they are 
proportional to the lepton mass, \ml. The different components of the
polarization \ie \pl, \pn, \pt involve different combinations of
Wilson coefficients and hence contain independent information. For
this reason confronting the polarization results with experiments are 
important investigations of the structure of SM and for establishing
new physics beyond it. The radiative process \bsllg has been
extensively studied in 2HDM and SUSY by various people
\cite{Iltan:2000iw,Xiong:2001up} and the importance of the neutral
Higgs bosons (NHBs) has been emphasized in the decay mode with $\mu$ and 
$\tau$ pairs in the final state. In this work we study  
various polarization asymmetries associated with final state lepton
(considering lepton to be either muon or tau) with special focus on
the NHB effects . 

\par \bsllg decay is induced by the pure leptonic decay \bsll which
suffers from helicity suppression for light leptons ($\ell = e,
\mu$). But in radiative mode (\bsllg) this helicity suppression is
overcome because the lepton pair by itself does not carry the
available four momentum . For this reason, one can expect \bsllg to 
have a relatively large branching ratio compared to non-radiative mode
despite an extra factor of $\alpha$. In MSSM, the situation for pure
dileptonic modes (\bsll) becomes different specially if $\ell = \mu,
\tau$ and \tanbeta is large  
\cite{Skiba:1993mg,Choudhury:1999ze,Xiong:2001up}. This is because in
MSSM the scalar and pseudoscalar Higgs coupling to the leptons is
proportional to $m_\ell tan\beta$ and thus can be large for $\ell =
\mu, \tau$ and for large \tanbeta. The effect of NHBs 
 has been studied in great detail in various leptonic decay
modes
\cite{Iltan:2000iw,Aliev:1999kr,Skiba:1993mg,Dai:1997vg,Choudhury:1999ze,Xiong:2001up,Huang:1999vb,RaiChoudhury:1999qb,Bobeth:2001jm,Grossman:1997qj}. 
The effect of NHBs on radiative mode \bsllg has also been studied in
2HDM \cite{Iltan:2000iw} and SUSY \cite{Xiong:2001up}. Here we will
focus on the NHB effects on various polarization asymmetries
within the framework of MSSM. 

\par This paper is organized as follows : In section 2, we first
present the Leading Order (LO) QCD corrected effective Hamiltonian 
for the quark level
process \btosllg including NHB effects leading to the
corresponding matrix element and
dileptonic invariant mass distribution. In section 3, 
all the three polarization asymmetries associated with the
final state lepton are calculated. Section 4 contains discussion of 
 the numerical
analysis of the polarization asymmetries and their dependence on
various parameters of the theory, focusing again mainly on NHB effects
in the large \tanbeta regime. 


\section{\label{section:2}Dilepton invariant mass distribution}

 The exclusive decay \bsllg can be obtained from the inclusive decay
\btosllg and further from \btosll. To do this photon has to be
attached to any charged internal or external line in the Feynman
diagrams for \btosll. As pointed out by Eilam \etal \cite{Eilam:1997vg},
contributions coming from attachment of photon to any charged
internal line will be suppressed by a factor of $m_b^2/M_W^2$ in
Wilson coefficient and hence can be safely neglected . So we only
consider the cases when the photon is hooked to initial quark 
lines and final lepton lines. To start off, the effective Hamiltonian
relevant for \btosll is 
\cite{Iltan:2000iw,Aliev:1999kr,Skiba:1993mg,Choudhury:1999ze,Xiong:2001up,Huang:1999vb,RaiChoudhury:1999qb}
:
\beqa
{\cal H}_{eff} &=& \frac{\alpha G_F}{2 \sqrt{2} \pi} V_{tb}V_{ts}^* 
             \left\{ ~ -2 ~ \c7eff ~\frac{m_b}{p^2} ~
                 \bar{s} i\sigma_{\mu\nu}p^\nu (1+\gamma_5) b
	             ~ \bar{\ell} \gamma^\mu \ell 
                  ~+~ \c9eff ~\bar{s}\gamma_\mu(1-\gamma_5) b
                   ~\bar{\ell}\gamma^\mu\ell  \right.    \nonumber \\
         &&   \left. ~+ ~C_{10} ~\bar{s}\gamma_\mu (1-\gamma_5) b
                      ~\bar{\ell}\gamma^\mu\gamma_5\ell
                    ~+~ \cq1 ~\bar{s} (1+\gamma_5) b
                        ~\bar{\ell}\ell
                    ~+~ \cq2 ~\bar{s}(1+\gamma_5)b
                        ~\bar{\ell}\gamma_5\ell ~
              \right\}
\label{eq:1}
\eeqa
where $p = p_1 + p_2$ is the sum of momenta of $\ell^-$ and $\ell^+$ and 
$V_{tb}, V_{ts}$ are CKM factors. The Wilson coefficients $\c7eff ,
\c9eff $ and $C_{10}$ are given in 
\cite{Grinstein:1989me,Cho:1996we}. Wilson coefficients
$\cq1 $ and $\cq2 $ are given in
\cite{Xiong:2001up,Huang:1999vb,Choudhury:1999ze,Bobeth:2001jm}. In
addition to the short 
distance corrections included in the Wilson coefficients, there are
some long distance effects also, associated with real $c \bar{c}$
resonances in the intermediate states. This is taken into account by
using the prescription given in \cite{Long-Distance}, namely by using
the Breit-Wigner form of resonances that add on to $\c9eff $ :
\beq
C_9^{(res)} ~=~
\frac{-3 \pi}{\alpha^2 } ~ \kappa_V
\sum_{V = J/\psi,\psi',..}
\frac{M_V Br(V \to l^+ l^-)\Gamma_{total}^V}{(s - M_V^2)
+ i \Gamma_{total}^V M_V} ~; 
\label{eq:2}
\eeq
there are six known resonances in $c \bar{c}$ system that can
contribute \footnote{all these six resonances will contribute to the
channel \bsmupmg whereas in the mode \bstaupmg  all but the lowest one
$J/\Psi(3097)$ will contribute because mass of this resonance is less
than the invariant mass of the lepton pair ($4 m_\ell^2$)}. The
phenomenological factor $\kappa_V$ is taken as 2.3 in
numerical calculations \cite{Kruger:1996cv,Long-Distance}.

\par Using eq(\ref{eq:1}) we calculate the matrix elements for the
decay mode \bsllg. When the photon is hooked to the initial
quark lines, the corresponding matrix element can be written as :
\beqa
{\cal M}_1 = \frac{\alpha^{3/2} G_F}{\sqrt{2 \pi}} V_{tb} V_{ts}^*
           &&  \left\{ ~[ A ~\epsmualbesig {\epsilon^*}^\alpha p^\beta
                  q^\sigma ~+~ i B ~
                 (\epsilon_\mu^*(pq)-(\epsilon^*p)q_\mu)] ~ 
                 \bar{\ell}\gamma^\mu\ell
               \right.                                 \nonumber\\
           && + ~ \left. [ C ~\epsmualbesig {\epsilon^*}^\alpha p^\beta
                   q^\sigma ~+~ i D ~
                   (\epsilon_\mu^*(pq)-(\epsilon^*p)q_\mu) ] ~
                   \bar{\ell}\gamma^\mu\gamma_5\ell   ~
                \right\}
\label{eq:3}
\eeqa
where A, B, C and D are related to the form factor definition and are
define in appendix eqns.(\ref{app:eq:1} - \ref{app:eq:4}). Here
$\epsilon_\mu$ and $q_\mu$ are the polarization vector and four
momentum of the photon respectively, $p$ is the momentum transfer to
the lepton pair \ie
the sum of momenta of $\ell^+$ and $\ell^-$. We can very easily see
from the structure of eq.(\ref{eq:3}) that neutral scalars don't  
contribute to ${\cal M}_1$ . This is due to eq.(\ref{app:eq:4}) given
in appendix.  

\par When the photon is radiated from either of the lepton lines we
get the contribution due to $C_{10}$ along with scalar and
pseudoscalar interactions \ie $\cq1 $ and $\cq2 $. 
Using eqns. (\ref{app:eq:5} - \ref{app:eq:7}) of the appendix
\cite{Iltan:2000iw,Xiong:2001up} the corresponding matrix element is :
\beqa
{\cal M}_2 
   &=&  \frac{\alpha^{3/2} G_F}{\sqrt{2\pi}} ~V_{tb} V_{ts}^* ~
       i 2 ~m_{\ell} ~
        f_{B_s} ~
        \left\{ ~( C_{10} ~+~ \frac{m_{B_s}^2}{2 m_{\ell} m_b} \cq2) ~
           \bar{\ell} \left[ \frac{\not\epsilon \not P_{B_s}}{2p_2q}
              ~-~ \frac{\not P_{B_s}\not\epsilon}{2p_1q}
                      \right] \gamma_5\ell
        \right.                              \nonumber        \\
   &&  ~+~ \left. \frac{m_{B_s}^2}{2 m_{\ell} m_b} \cq1 
              \left[ 2 m_\ell ( \frac{1}{2p_1q} ~+~ \frac{1}{2p_2q}) ~
                  \bar{\ell}\not\epsilon\ell
        ~+~ \bar{\ell}( \frac{\not\epsilon \not P_{B_s}}{2p_2q}
        ~-~ \frac{\not P_{B_s}\not\epsilon}{2p_1q}) \ell
          \right]
        \right\}.
\label{eq:4}
\eeqa
where $P_{B_s}$ and $f_{B_s}$ are the four momentum and decay constant
of $B_s$ meson and  $p_1$ and  $p_2$ are the four momenta of $\ell^-$ and
$\ell^+$ respectively.  

\par The final matrix elment of \bsllg decay thus is :
\beq
{\cal M} ~=~ {\cal M}_1 ~+~ {\cal M}_2
\label{eq:5}
\eeq
From this matrix element we can get the square of the matrix element
as :
\beq
\modmsq ~=~ \modm1sq ~+~ \modm2sq ~+~ 2 \rem12
\label{eq:6}
\eeq
with
\beqa
\modm1sq &=& 4 ~ \facmatrix ~ 
             \left\{ 
            ~[ ~\modsq{A} + \modsq{B}~ ] ~ [ \psq ( \p1qsq + \p2qsq ) 
               + 2 m_\ell^2 \pqsq ] ~+~ [~ \modsq{C} + \modsq{D}~ ] 
	     \right.                  \nonumber   \\
         &&  \left. ~ [ \psq ( \p1qsq +  \p2qsq ) - 2 m_\ell^2 \pqsq ]
                ~+~ 2 ~Re(B^* C + A^*D)~ \psq ( \p2qsq - \p1qsq ) 
             \right\}    
\label{eq:7}             \\
\modm2sq &=& 
   4 ~ \facmatrix ~ f_{B_s}^2 ~ \mle2 ~
       \Bigg[
	  \ctencqtwo 
           \left\{ ~ 8 + 
              \frac{1}{\p1qsq } ( - 2 \mb2 \mle2 - \mb2 \psq + p^4 ~+~
                 2 p^2 ( p_2 q ) )
           \right.                    \nonumber  \\
 &&        \left.  
         .  ~+~  \frac{1}{(p_1 q)} ( 6 \psq + 4 (p_2 q) )
         ~+~  \frac{1}{\p2qsq } 
           ( - 2 \mb2 \mle2 - \mb2 \psq + p^4 + 2 \psq (p_1 q) )
         ~+~  \frac{1}{(p_2 q)} ( 6 \psq + 4 (p_1 q) )
           \right.                      \nonumber   \\
 &&        \left. 
         ~+~ \frac{1}{(p_1 q) (p_2 q)} ( - 4 \mb2 \mle2 + 2 p^4 )
            ~ \right\}                  \nonumber    \\
 &&      +~  \drcqone 
         \left\{ 8  ~+~  \frac{1}{\p1qsq } 
         . ( 6 \mb2 \mle2 + 8 \mle4 - \mb2 \psq - 8 \mle2 \psq 
            + p^4 - 8 \mle2 (p_2 q) + 2 \psq (p_2 q) ) 
         \right.                      \nonumber    \\
 &&      \left. 
             +~ \frac{1}{ (p_1 q) } (- 40 \mle2 + 6 \psq + 4 (p_2 q) )
             ~+~  \frac{1}{\p2qsq } ( 6 \mb2 \mle2 + 8 \mle4 - \mb2
           \psq - 8 \mle2 \psq + p^4 - 8 \mle2 (p_1 q) 
         \right.                      \nonumber   \\
 &&      \left.  +~ 2 \psq (p_1 q) )
             ~+~ \frac{1}{ (p_2 q) } ( - 40 \mle2 + 6
              \psq  + 4 (p_2 q) )
        +~ \frac{1}{(p_1 q) (p_2 q)} ( 4 \mb2 \mle2 + 16 \mle4
                 - 16 \mle2 p^2  + 2 p^4 ) 
         \right\}
	\Bigg]                                    
\label{eq:8}              
\eeqa
\beqa
2 Re( {\cal M}_1 {\cal M}_2^* ) 
 &=&  
16 ~ \facmatrix ~ f_{B_s} ~ \mle2 
       \Bigg[  \ctencqtwo 
               \left\{ ~ - ~ Re(A) \frac{( p_1 q + p_2 q)^3}{(p_1 q)
                      (p_2q)}
               \right.               \nonumber    \\
 &&            \left. +~ Re(D) \frac{ (p q)^2 ( p_1 q ~-~ p_2 q ) }
                        {(p_1 q) (p_2 q)} 
               \right\} 
               ~+~ \drcqone ~ 
               \left\{ Re(B) ~ \frac{1}{(p_1 q)(p_2 q)} ( - (p q)^3 
	       - 2 (p_1 p_2) (p_1 q)^2 
               \right.               \nonumber    \\
 &&            \left. ( - 2 (p_1 p_2) (p_2 q)^2 
                + 4 \mle2 (p_1 q)(p_2q)) 
               ~+~ Re(C) \frac{(p q)^2 ( p_1 q - p_2 q )}{(p_1 q)
                   (p_2 q)} 
               \right\}
\label{eq:9}
\eeqa
The differential decay rate of \bsllg as a function of invariant mass
of dileptons is given by :
\beqa
\frac{d \Gamma}{d\sh} 
   = ~\facdr ~\frac{\mb5}{16 (2 \pi)^3}~ (1 - \sh) ~ \faco
      ~ \bigtriangleup 
\label{eq:10}
\eeqa
with $\bigtriangleup$ defined as
\beqa
\bigtriangleup &=& ~~ 
      {4 \over 3} ~\mb2 ~( 1 - \sh)^2~ 
        [ ~ ( \modsq{A} + \modsq{B} )~ (2 \mlhsq + \sh) 
          ~+~ ( \modsq{C} + \modsq{D} ) ( - 4 \mlhsq + \sh) ~ ]
                                                     \nonumber \\
  &&  +~ \frac{64 ~f_{B_s}^2 \mlhsq}{\mbs^2} ~\ctencqtwo^2 ~ \frac{[~ (1 - 4 
          \mlhsq + \sh^2) \lnzh ~-~ 2 \sh \faco ~ ] }{(1 -
          \sh)^2 ~ \faco}                            \nonumber \\
  &&  -~ \frac{64 ~f_{B_s}^2 \mlhsq}{\mbs^2} ~ \drcqone^2 ~ 
         \frac{[~ ( - 1 + 12 \mlhsq - 16 \mlh^4 - \sh^2)
              \lnzh ~+~ ( -2 \sh - 8 \mlhsq \sh + 4 \sh^2 ) \faco ~]}
              {(1 - \sh)^2 ~ \faco }        \nonumber \\ 
  &&  +~ 32 ~ f_{B_s} \mlhsq ~\ctencqtwo ~ Re(A) ~\frac{\lnzh}{\faco} 
                              \nonumber  \\
  &&  
      -~ 32 ~ f_{B_s} \mlhsq ~\drcqone ~Re(B)~ \frac{[ ~ (1 - 4
          \mlhsq + \sh) \lnzh ~-~ 2 \sh \faco ~]}{\faco}
\eeqa
where $\sh = \psq/\mb2 ~,~ \mlhsq = m_\ell^2/\mb2 ~,~ \zh = \frac{1 + 
\faco}{1 - \faco}$ are dimensionless quantities


\section{\label{section:3}Lepton Polarization asymmetries}

We now compute the lepton polarization asymmetries from the four Fermi
interaction defined in the matrix element eqn.(\ref{eq:3}) and
eqn.(\ref{eq:4}). 
For this we need to calculate the polarized rates corresponding to
different lepton polarizations. These rates are obtained by
introducing spin projection operators defined by $N ~=~ {1 \over 2} (1
+ \gamma_5 {\not S}_x)$, where index $x ~=~ L, ~N, ~T$ and corresponds
to longitudinal, normal and transverse polarization states
respectively. The orthogonal unit vectors, $S_x$, defined in the rest
frame of $\ell^-$ read \cite{Kruger:1996cv} : 
\beqa
S^\mu_L &\equiv& (0, \ev{L}) ~=~ \left(0, \frac{\pv}{|\pv|}\right)
\nonumber \\ 
S^\mu_N &\equiv& (0, \ev{N}) ~=~ \left(0, \frac{\qv \times \pv}{|\qv
        \times  \pv|} \right)         \nonumber   \\
S^\mu_T &\equiv& (0, \ev{T}) ~=~ (0, \ev{N} \times \ev{L}) 
\label{eq:11}
\eeqa
where $\pv $ and $\qv $ are the three momenta of $\ell^-$ and photon
in the center-of-mass (CM) frame of $\ell^- \ell^+$
system. Furthurmore, it is quite obvious to note that, $S_x . p_1 ~=~
0$.
 Now
boosting all the three vectors given in eqn.(\ref{eq:11}) to the
dilepton rest frame , only the longitudinal vector will get boosted
while the other two (normal and transverse) will remain the same. The 
longitudinal vector after boost  becomes \footnote{this particular
choice of polarization is called helicity} : 
\beq
S^\mu_L ~=~ \left( \frac{|\pv|}{m_\ell}, \frac{E_1 \pv}{m_\ell |\pv|} 
\right)
\label{eq:12}
\eeq
We can now calculate the polarization asymmetries by using the spin
projectors for $\ell^-$ as ${1 \over 2} ( 1 + \gamma_5 \not S)$. The
lepton polarization asymmetries are defined as :
\beq
P_x(\sh) ~~~\equiv~~~
               \frac{ \frac{d \Gamma(S_x)}{d \sh}  ~-~ \frac{d \Gamma(-
                      S_x)}{d \sh}}{\frac{d \Gamma(S_x)}{d \sh}  
                      ~+~ \frac{d \Gamma(- S_x)}{d \sh}}
\label{eq:13}
\eeq
where the index $x$ is $L, T$ or $N$, representing respectively the
longitudinal asymmetry, the asymmetry in the decay plane and the
normal component to the decay plane. From the definition of the
lepton polarization we can see that \pl and \pt are P-odd, T-even and
CP-even observable while \pn is P-even, T-odd and hence CP-odd
observable \footnote{because time reversal operation changes the signs
of momentum and spin, and parity transformation changes only the sign
of momentum}

\par Our results for the polarization asymmetries are 
\beqa
P_L(\sh) &=& 
     \Bigg[
         {8 \over 3} \mb2 ~Re( A^* C + B^* D )~ \faco 
        ~\sh ( 1 -  \sh )^2
      -~ \frac{128 f_{B_s}^2 \mlhsq}{\mb2 }~ \ctencqtwo 
                                   \nonumber \\
  &&      \times \drcqone \frac{1}{(1 - \sh)^2 (\sh - 4 \mlhsq )}
              \Bigg\{  ( \sh - 4 \mlhsq \sh - 2 \sh^2 
               - 4 \mlhsq \sh^2 + 3 \sh^3)\faco  
                           \nonumber  \\
  &&
                 +~ ( 2 \mlhsq - 8 \mlh^4 - \sh + 8 \mlhsq \sh 
                 - 8 \mlh^4 \sh + 2 \mlhsq \sh^2 - \sh^3)
               ~\lnzh 
            \Bigg\}       \nonumber   \\
  && +~   32 ~ f_{B_s} \mlhsq ~\ctencqtwo  \frac{1}{(4 \mlhsq - \sh)}
          \left\{ Re(B) \Bigg( (- \sh + 3 \sh^2) \faco ~+~ 2 ( \mlhsq
               + \mlhsq \sh - \sh^2) \lnzh \Bigg)
          \right.                  \nonumber    \\
  &&      \left. -~ Re(C) (1 - \sh) \Bigg( \sh \faco
            + (2  \mlhsq - \sh) \lnzh \Bigg)
          \right\}        
           ~+~ 32 f_{B_s} \drcqone \frac{\mlhsq (1 - \sh)}{\sh \fact} 
                                   \nonumber    \\
  &&     \left\{  Re(A) ~\Bigg( - \sh ~ \faco ~+~ 2 \mlhsq ~\lnzh
             ~\Bigg) 
           +~ Re(D) \Bigg( \sh \faco + (2 \mlhsq - \sh) \lnzh \Bigg)
         \right\}
             \Bigg]/\bigtriangleup                     
\label{eq:14}       \\  
P_T(\sh) &=& \pi ~ \mlh 
  \Bigg[
      -~  2 \mb2  ~Re(A^* B) ~ \sqrt{\sh} (1 - \sh)^2  
      ~-~ \frac{64 f_{B_s}^2}{\mb2 } \mlh \ctencqtwo \drcqone 
         \frac{(1 - 4 \mlhsq)}{(1 - \sh)} \nonumber   \\
   && +~  8 ~f_{B_s} ~\ctencqtwo 
          \left\{  Re(B)~ \frac{(1 - \sh)( \sh + 4 \mlhsq )}{(2 \mlh +
         \sqrt{\sh})}
                ~+~ Re(C) ~ ( - 2 \mlh + \sqrt{\sh})~(1 + \sh) 
          \right\}                  \nonumber   \\
   && +~  8 ~f_{B_s} ~ \drcqone 
          \left\{  Re(A) \frac{(4 \mlhsq + \sh - 12 \mlhsq \sh +
               \sh^2)}{(2 \mlh + \sqrt{\sh})}
           ~-~ Re(D) ~ ( 2 \mlh - \sqrt{\sh}) ~ (1 - \sh)  
          \right\}           
     \Bigg]/\bigtriangleup
\label{eq:15}      \\   
P_N(\sh) &=& \pi ~ \mlh  ~
   \Bigg[
      - ~ \mb2 Im(A^* D  +  B^* C) ~(1 - \sh)^2 ~\sqrt{\sh - 4
           \mlhsq}                  \nonumber   \\
   && + ~  8 ~ f_{B_s} ~\ctencqtwo ~ \frac{\sh \faco}{(2 \mlh + \sqrt{s})}  
           \left\{ Im(A) (1 + \sh) ~+~ Im(D) (1 - \sh)
           \right\}                  \nonumber  \\
   && +~  8 ~f_{B_s} ~\drcqone ~ \frac{\sqrt{\sh - 4 \mlhsq}}{(2 \mlh +
            \sqrt{\sh})} 
           \left\{ Im(B) (1 - \sh) + Im(C) (1 - 8 \mlhsq + \sh) 
           \right\}
   \Bigg]/\bigtriangleup
\label{eq:16}
\eeqa
with $\bigtriangleup$ as defined in eqn.(\ref{eq:11}) and $\mlh =
m_\ell/m_{B_s}$. 


\section{\label{section:4}Numerical results and discussion}

     We have performed the numerical analysis of various
polarization asymmetries whose analytical expressions are given in 
eqns.(\ref{eq:14} - \ref{eq:16}).

\par Although MSSM is the simplest (and the one having the least
number of parameters) SUSY model, it still has a very large number of
parameters making it rather difficult to do any meaningfull
phenomenology in such a large parameter space. Many choices are
available to reduce such  
large number of parameters. The most favorite among them is the
Supergravity (SUGRA) model. In this model, universality of all the
masses and couplings is assumed at the GUT scale. The minimal SUGRA
(mSUGRA) model has only five parameters (in addition to SM
parameters) to deal with. They are : $m$ (the unified mass of all
the scalars), $M$ (unified mass of all the gauginos), \tanbeta (ratio
of vacuum expectation values of the two Higgs doublets), $A$ (the
universal trilinear coupling constant) and finally, $sgn(\mu)$
\footnote{our convention of the $sgn(\mu)$ is that $\mu$ enters the 
chargino mass matrix with positive sign} .

\par It has been well emphasized in many works 
\cite{Choudhury:1999ze,RaiChoudhury:1999qb,goto1} that it is not
necessary to have a common mass for all the scalars at the GUT
scale. To have 
required suppression in $K^0 - \bar{K^0}$ mixing, it is sufficient to
have common masses of all the squarks at the GUT scale. So the condition
of universality of all scalar masses at the GUT is not a very strict
one in SUGRA. 
Thus we also explore a more relaxed kind of mSUGRA model where
the condition of universality of all the scalar masses at the GUT
scale is relaxed with the assumption that  universal squark and Higgs
masses are  
different. For the Higgs sector we take the pseudo-scalar Higgs mass
($m_A$) to be a parameter. Over the whole MSSM parameter space we  
have imposed a 95 \% CL bound \cite{Anikeev:2002a}, consistant
with CLEO and ALEPH results :
$$ 2 \times 10^{-4} < Br(B \to X_s \gamma) < 4.5 \times 10^{-5}$$ 

\par Figure(\ref{fig:brmuon}) shows plots of the differential Braching
ratios of \bsllg for leptons to be $\mu$ and $\tau$ . The prediction
of the Branching ratios for 
\bsllg are : 
\begin{table}[h]
\caption{\label{table:1} Branching ratios for \bsllg ($\ell \ = \mu ,
\tau$)}  
\bcen
\begin{tabular}{| c | c  c | }  \hline
 \hspace{1cm} Model \hspace{1cm}  
     &   \hspace{.5cm} Br(\bsmupmg) \hspace{.5cm}  
     &  \hspace{.5cm} Br(\bstaupmg)  \hspace{.5cm} \\  \hline \hline 
 Standard Model  & $5.53 \times 10^{-8}$ & $6.57 \times 10^{-8}$  \\
 mSUGRA \footnotemark[1]  
         & $6.86 \times 10^{-8}$ & $3.59 \times 10^{-7}$  \\
 SUGRA \footnotemark[1] 
          & $1.21 \times 10^{-7}$ & $1.31 \times 10^{-6}$  \\ \hline
\end{tabular}
\ecen
\footnotetext[1]{ The mSUGRA and SUGRA parameters are defined in
Figure(\ref{fig:brmuon}). These values are of the same order as
estimated by Xiong \etal \cite{Xiong:2001up} }
\end{table}

We have plotted various polarization 
asymmetries (\pl , \pt and \pn) in the three models - SM, mSUGRA and
SUGRA in Figures(\ref{fig:longs},\ref{fig:norms},\ref{fig:transs})
 for \bsmupmg and \bstaupmg as a function of $\sh$ (scaled invariant
mass of the dilepton pair).

\par Now we try to analyse the behavior of the polarization
asymmetries on the parameters of the models chosen (mSUGRA, and
SUGRA). For this analysis we consider the polarization 
asymmetries at dilepton invariant mass ($\sh$) away from the
resonances (the $J/\Psi$) resonances (we choose $\sh = 0.68$ for our
analysis) . The main focus of the analysis is NHB effects on
polarization asymmetries. These effects crucially depend on \tanbeta 
and pseudoscalar Higgs mass ($m_A$). 

\par In mSUGRA model the Higgs mass (at electroweak scale) depends
crucially on  the universal mass of the scalars and \tanbeta . To
illustrate this crucial behaviour,   
we have plotted various polarization asymmetries as a function of
\tanbeta for different values of unified scalar mass ($m$) in
Figs. (\ref{fig:longtb}, \ref{fig:normtb}, \ref{fig:transtb}) .
As can be seen from these figures, \pl shows large
deviations from the SM values and over a significant portion of the
allowed region, even shows a sign flip, provided \tanbeta is
sufficiently large. Similar behaviour is also there for \pt. On the  
other hand, the predictions for \pn don't differ substantially from SM
results but the mSUGRA predictions can change \pn by more than 50 \%
with an appreciable increase in \tanbeta.  

\par For SUGRA model we have plotted ( Figs. (\ref{fig:longma},
\ref{fig:normma}, \ref{fig:transma})) the polarization asymmetries as
a  function of pseudoscalar Higgs mass ($m_A$) for various values of 
\tanbeta. In SUGRA we expect more variation of all the
polarization asymmetries as compared to their SM values because here
we have Higgs mass (pseudo-scalar Higgs mass) as an additional
parameter along with \tanbeta. As we can see from
Figure(\ref{fig:longma}) the variation of \pl is more substantial in
SUGRA model. In fact for fairly large region of SUGRA parameter space,
\pl can be opposite in sign as compared to SM case . \pt can vary upto
five in magnitude when compared with the SM value over the large
region of allowed parameter space and for the parameter 
space we have taken into consideration, the predicted value of \pt in
SUGRA is opposite in sign to the SM value. Again \pn does not show as
much deviation as observed for \pl and \pt but the variation can still
be upto an order in certain region.

\par Summarizing the results of the numerical analysis  :
\begin{enumerate}
\item{} From Figures (\ref{fig:longs},\ref{fig:transs}) it is  clear
that the longitudinal and transverse polarization asymmetries (\pl,
\pt) can have  substantial deviation from their respective Standard
Model values over the whole region of dilepton invariant mass ($\sh$),
while Figure(\ref{fig:norms}) indicates deviation for \pn from SM
values for a limited region of the dilepton invariant mass. 
 
\item{} As we have pointed out earlier \cite{RaiChoudhury:1999qb} that
for the inclusive process $B \to X_s \ell^+ \ell^-$ there is not much
deviation from SM results in mSUGRA model. But in radiative dileptonic
decay mode, mSUGRA predictions also show large deviations (at least of
\pl and \pt) from SM results, making it possible to use polarization
asymmetries to test mSUGRA model. This is mainly because in the
bremmstrahlung part of the matrix elment (${\cal M}_2$), the Wilson
coefficient $\cq2 $ adds on to $C_{10}$ via the combination
$\ctencqtwo$ which effectively increases the SM value of
$C_{10}$. This doesn't happen for the process $B \to X_s \ell^+
\ell^-$ and this numerically is the reason for the scalar exchanges
affecting the \bsllg process more than the semi-leptonic one. 

\item{} From Figs. (\ref{fig:longtb}, \ref{fig:normtb},
\ref{fig:transtb}) we can see that the polarization asymmetries show a
general enhancement with increase in \tanbeta and they decrease as the
universal scalar mass ($m$) is increased. This is expected because the
Higgs boson mass increases with $m$ and thus the contributions of
scalar ($\cq1 $) and pseudoscalar ($\cq2 $) type interactions
decrease. 

\item{} As can be  seen from the structure of the analytical
expressions for various polarization asymmetries ( eqn.(\ref{eq:14},
\ref{eq:15}, \ref{eq:16}), they are all different analytic functions
of various Wilson coefficients and hence contain independent
information. These asymmetries, hence, can also be used for accurate
determination of various Wilson coefficients.  
\end{enumerate}

\par In conclusion, we can say that the observation of the polarization
asymmetries can be a very useful probe for finding out the new physics
effects and testing the structure of the effective Hamiltonian. 

\begin{acknowledgments}
NM would like to thank the University Grants Commission, India for the
financial support.
\end{acknowledgments}


\appendix


\section{\label{inpara} Input Parameters} 

\bcen
$\mbs ~=~ 5.26$ GeV \ , \  $m_c ~=~ 1.4$ GeV  \ , \  $m_s ~=~ 0.2$ GeV
\\
$m_\mu ~=~ 0.106$ GeV \ , \ $m_\tau ~=~ 1.77$ GeV \ , \ $m_b ~=~ 4.8$
GeV \\
$m_W ~=~ 80.4$ GeV \ , \  $m_t ~=~ 176$ GeV  \ , \ $|V_{tb} V_{ts}^*|
~=~ 0.045$  \\
$G_F ~=~ 1.17 \times 10^{-5} ~{\rm GeV}^{-2}$ \ , \ $\alpha ~=~
\frac{1}{129} $  \ , \ $\tau(\mbs) ~=~ 1.6 \times 10^{-12}$ sec
\ecen


\section{\label{section:5}}

Definition of A, B, C and D defined in eq(\ref{eq:3}) are :
\beqa
A  &=&  \frac{1}{m_{B_s}^2} ~[ \c9eff G_1(p^2) ~-~  2 \c7eff
        \frac{m_b}{p^2}G_2(p^2)],                   \nonumber  \\ 
B  &=&  \frac{1}{m_{B_s}^2}~ [ \c9eff F_1(p^2) ~-~ 2 \c7eff
        \frac{m_b}{p^2}F_2(p^2)],                   \nonumber  \\ 
C  &=&  \frac{C_{10}}{m_{B_s}^2}~G_1(p^2),           \nonumber  \\
D  &=&  \frac{C_{10}}{m_{B_s}^2}~F_1(p^2).
\label{app:eq:1}
\eeqa
where the form factors definition chosen is
\cite{Eilam:1995zv} 
\beqa
\langle\gamma |~ \bar{s} \gamma_\mu (1 \pm \gamma_5) b ~|B_s \rangle
   &=&   \frac{e}{m_{B_s}^2}
         \left\{ \epsmualbesig \epsilon_\alpha^* p_\beta q_\sigma 
             G_1(p^2)\mp i [ (\epsilon_\mu^*(pq)-(\epsilon^*p)q_\mu) ]
            F_1(p^2) 
         \right\}                                          
\label{app:eq:2}       \\
\langle\gamma| ~\bar{s} i \sigma_{\mu\nu} p_\nu (1 \pm \gamma_5) b ~
 |B_s\rangle 
  &=&   \frac{e}{m_{B_s}^2}
        \left\{  \epsmualbesig \epsilon_\alpha^* p_\beta q_\sigma 
           G_2(p^2) \pm i [ (\epsilon_\mu^*(pq)-(\epsilon^*p)q_\mu) ]
           F_2(p^2) 
        \right\}
\label{app:eq:3}     
\eeqa
multiplying eq.(\ref{app:eq:2}) with $p_\mu$ and using equation of
motion we can get relation :
\beq
\langle\gamma|~ \bar{s} (1 \pm \gamma_5) b ~|B_s\rangle ~=~ 0
\label{app:eq:4}
\eeq
The defination of form factors we are using for numerical analysis is
\cite{Eilam:1995zv} : 
\beqa
G_1(\psq) ~=~ \frac{1}{1 - \psq/5.6^2} ~GeV &,& G_2(\psq) ~=~
\frac{3.74}{1 - \psq/40.5} GeV^2 ,        \nonumber   \\
F_1(\psq) ~=~ \frac{0.8}{1 - \psq/6.5^2} ~ GeV &,& F_2(\psq) ~=~
\frac{0.68}{1 - \psq/30}~GeV^2 . 
\label{app:eq:4a}
\eeqa

Identities used in calculation of matrix element when photon is
radiated from lepton leg : 
\beqa
\langle 0|~\bar{s} b ~|B_s\rangle &=& 0 
\label{app:eq:5}           \\
\langle 0|~\bar{s}\sigma_{\mu\nu} (1 + \gamma_5) b~|B_s\rangle &=& 0 
\label{app:eq:6}                        \\
\langle 0|~ \bar{s} \gamma_\mu \gamma_5 b ~|B_s\rangle &=& -~ i
f_{B_s} P_{B_s\mu}
\label{app:eq:7}
\eeqa




\pagebreak 


\bfig
 \vskip 1.2cm
 \bcen
    \epsfig{file=brmuons.eps,height=3.7in}
    \vskip 1.4cm
    \epsfig{file=brtaus.eps,height=3.7in}
\caption{Branching ratios for \bsllg with $\ell ~=~ \mu$ (above) and
$\ell ~=~ \tau$ (below). mSUGRA parameters are : $m = 200$ GeV, $M =
450$ GeV , $A = 0$ , $\tan\beta = 40$. Additinal parameter for SUGRA
(the pseudo-scalar Higgs mass) is taken to be $m_A = 306$ GeV}
\label{fig:brmuon}
\ecen
\efig


\bfig
 \vskip 1.2cm
 \bcen
    \epsfig{file=pl_muon_s.eps,height=3.7in}
    \vskip 1.4cm
    \epsfig{file=pl_tau_s.eps,height=3.7in}
\caption{$P_L$ for \bsllg with $\ell ~=~ \mu$ (above) and
$\ell ~=~ \tau$ (below). mSUGRA parameters are : $m = 200$ GeV, $M =
450$ GeV , $A = 0$ , $\tan\beta = 40$. Additinal parameter for SUGRA
(the pseudo-scalar Higgs mass) is taken to be $m_A = 306$ GeV}
\label{fig:longs}
\ecen
\efig


\bfig
 \vskip 1.3cm
  \bcen
     \epsfig{file=pl_tb.eps,height=3.6in}
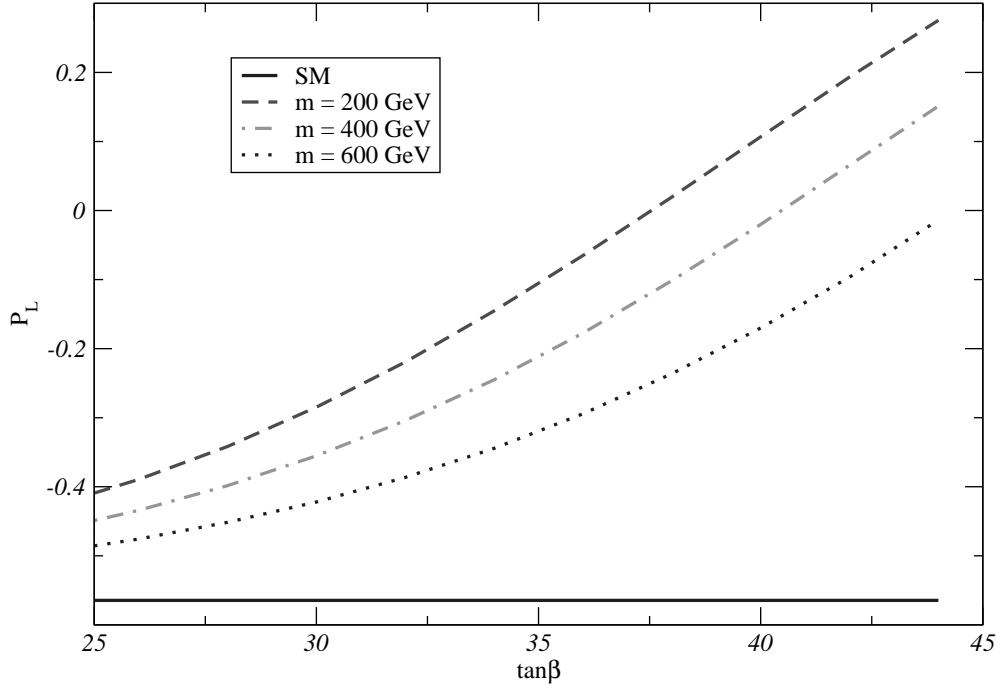
\caption{\pl vs \tanbeta at $\sh = 0.68$ for \bstaupmg in mSUGRA
model, other parameters are : $M = 450$ GeV , $A = 0$.} 
  \label{fig:longtb}
  \vskip 1.7cm
     \epsfig{file=pl_ma.eps,height=3.6in}
\caption{\pl vs $m_A$ at $\sh = 0.68$ for \bstaupmg in SUGRA, other
parametes are : $m = 200$ GeV , $M = 450$ GeV , $A = 0$.} 
  \label{fig:longma}
  \ecen
\efig


\bfig
 \vskip 1.2cm
 \bcen
    \epsfig{file=pn_muon_s.eps,height=3.7in}
    \vskip 1.4cm
    \epsfig{file=pn_tau_s.eps,height=3.7in}
\caption{\pn for \bsllg with $\ell ~=~ \mu$ (above) and
$\ell ~=~ \tau$ (below). mSUGRA parameters are : $m = 200$ GeV, $M =
450$ GeV , $A = 0$ , $\tan\beta = 40$. Additinal parameter for SUGRA
(the pseudo-scalar Higgs mass) is taken to be $m_A = 306$ GeV}
\label{fig:norms}
\ecen
\efig


\bfig
  \vskip 1.3cm
  \bcen
     \epsfig{file=pn_tb.eps,height=3.6in}
\caption{\pn vs \tanbeta for \bstaupmg at $\sh = 0.68$ in mSUGRA ,
other parameters are : $M = 450$ GeV , $A = 0$.}  
  \label{fig:normtb}
  \vskip 1.5cm
     \epsfig{file=pn_ma.eps,height=3.6in}
\caption{\pn vs $m_A$ at $\sh = 0.68$ for \bstaupmg in SUGRA , other
parametes are : $m = 200$ GeV , $M = 450$ GeV , $A = 0$.} 
  \label{fig:normma}
  \ecen
\efig


\bfig
 \vskip 1.2cm
 \bcen
    \epsfig{file=pt_muon_s.eps,height=3.7in}
    \vskip 1.4cm
    \epsfig{file=pt_tau_s.eps,height=3.7in}
\caption{\pt for \bsllg with $\ell ~=~ \mu$ (above) and
$\ell ~=~ \tau$ (below). mSUGRA parameters are : $m = 200$ GeV, $M =
450$ GeV , $A = 0$ , $\tan\beta = 40$. Additinal parameter for SUGRA
(the pseudo-scalar Higgs mass) is taken to be $m_A = 306$ GeV}
\label{fig:transs}
\ecen
\efig


\bfig
  \bcen
     \epsfig{file=pt_tb.eps,height=3.7in}
\caption{\pt vs \tanbeta for \bstaupmg at $\sh = 0.68$ in mSUGRA ,
other parameters are : $M = 450$ GeV , $A = 0$.}  
  \label{fig:transtb}
  \vskip 1.5cm
     \epsfig{file=pt_ma.eps,height=3.7in}
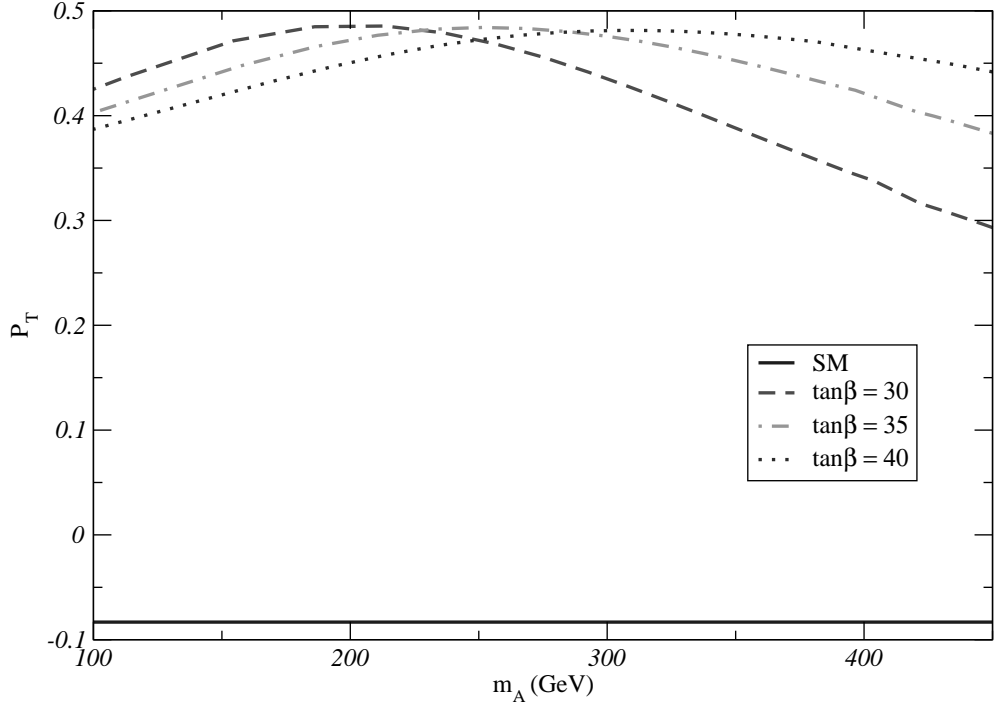
\caption{\pt vs $m_A$ for \bstaupmg at $\sh = 0.68$ in SUGRA , other
parameters are : $m = 200$ GeV , $M = 450$ GeV , $A = 0$.} 
  \label{fig:transma}
  \ecen
\efig


\end{document}